\documentclass[11pt,a4paper]{article}
\usepackage{amsmath}
\usepackage[mathcal]{eucal}
\usepackage{latexsym}
\usepackage{amssymb}
\newcommand*{\eps}{{\rlap{\lower2ex\hbox{$\,\,\tilde{}$}}{\epsilon_{ijk}}}}
\newcommand*{\EPS}{{\rlap{\lower2ex\hbox{$\,\,\tilde{}$}}{\epsilon_{i'j'k'}}}}
\newcommand*{\lmq}{{\rlap{\lower2ex\hbox{$\,\,\tilde{}$}}{\epsilon_{lmq}}}}
\newcommand*{\jmq}{{\rlap{\lower2ex\hbox{$\,\,\tilde{}$}}{\epsilon_{jmq}}}}
\newcommand*{\jql}{{\rlap{\lower2ex\hbox{$\,\,\tilde{}$}}{\epsilon_{jql}}}}
\newcommand*{\jlm}{{\rlap{\lower2ex\hbox{$\,\,\tilde{}$}}{\epsilon_{jlm}}}}
\newcommand*{\imq}{{\rlap{\lower2ex\hbox{$\,\,\tilde{}$}}{\epsilon_{imq}}}}
\newcommand*{\iql}{{\rlap{\lower2ex\hbox{$\,\,\tilde{}$}}{\epsilon_{iql}}}}
\newcommand*{\ilm}{{\rlap{\lower2ex\hbox{$\,\,\tilde{}$}}{\epsilon_{ilm}}}}
\newcommand*{\lmn}{{\rlap{\lower2ex\hbox{$\,\,\tilde{}$}}{\epsilon_{lmn}}}}
\newcommand*{\abc}{{\rlap{\lower2ex\hbox{$\,\,\tilde{}$}}{\epsilon_{abc}}}}
\newcommand*{\N}{{\rlap{\lower2ex\hbox{$\,\,\tilde{}$}}{N}}}
\newcommand{\tN}{{\rlap{\lower2ex\hbox{$\,\,\tilde{}$}}{N}}}
\newcommand*{\tM}{{\rlap{\lower2ex\hbox{$\,\,\tilde{}$}}{M}}}
\begin{document} 
\begin{titlepage} 
\title{Reformulation of the Hamiltonian constraint of four dimensional gravity for arbitrary values of the Immirzi parameter via the affine group formalism.}
\medskip
\author{Eyo Eyo Ita}
\medskip  
\input amssym.def
\input amssym.tex
\maketitle
\centerline{\it $^{2}$Department of Physics, US Naval Academy, Annapolis Maryland} 
\smallskip
          
\bigskip    
                
\begin{abstract}
One of the virtues of the Ashtekar variables is the simplification of the initial value constraints for gravity.  In the case of self-dual variables this entails a complexification of the phase space which comes at the expense of having to implement reality conditions in the Lorentzian signature case.  A reformulation of the theory in terms of real variables eliminates this difficulty, albeit at the expense of having to deal with a more complicated Hamiltonian constraint.  The set of available gravitational theories classically equivalent to Einstein's is parametrized by a parameter $\beta$, known as the Immirzi parameter.  We rephrase the Hamiltonian constraint into the form of an affine Lie algebra for arbitrary $\beta$, and perform a quantization.
\end{abstract}
\end{titlepage}

\section{Introduction}

Consistent quantization of Lorentzian signature, four-dimensional General Relativity (GR) remains one of the most challenging problems in theoretical physics.  Part of the main difficulties can be traced back to the implementation and interpretation of the Hamiltonian constraint in the quantum theory.  The initial attempts were made within the framework of metric variables \cite{DEWITT}, where it became clear that the nonpolynomial nature of the constraint, also known as the Wheeler--DeWitt equation, rendered it seemingly intractable.  In the mid 1980's it was discovered by Abhay Ashtekar that GR could be written based on the phase space of a Yang--Mills theory using new variables \cite{ASH}, \cite{ASH1}.  The initial value constraints, including the Wheeler--DeWitt equation, were written as low order polynomials on the phase space, which opened up new avenues toward the addressal of a background-independent, nonperturbative quantum theory of gravity.  The basic variables were a self-dual $SU(2)$ connection $A^a_i=\Gamma^a_i\pm{i}K^a_i$ and a densitized triad $\widetilde{E}^i_a=\frac{1}{2}\widetilde{\epsilon}^{ijk}\epsilon_{abc}e^b_je^c_k$.\footnote{Index conventions in this paper are such that internal $SU(2)$ indices are denoted by $a,b,c,\dots$, while spatial indices are denoted by $i,j,k,\dots$, each taking values $1-3$.  The spacetime manifold $M$ is foliated into spatial slices $\Sigma$, which are labelled by the parameter $t$.}  Here, $K^a_i$ is the triadic form of the extrinsic curvature $K_{ij}=K^a_ie^a_j$ of 3-space $\Sigma$, while $\Gamma^a_i=-\frac{1}{2}\epsilon^{abc}\omega^{bc}_i$ is the Levi--Civita connection compatible with the triads $e^a_i$. The simplification of the constraints in the Ashtekar variables entails complexification of the gravitational phase space.  This complexification comes at a price of having to satisfy reality conditions in order to ensure that in the end, one indeed has real GR of Lorentzian signature, conditions which in the quantum theory have proven to be nontrivial to implement.\par 
\indent 
It was shown by Barbero \cite{BARBERO}, \cite{BARBERO1} that under a certain canonical transformation one could formulate the Lorentzian signature theory based on a real $SU(2)$ connection $A^a_i=\Gamma^a_i\pm{K}^a_i$, thus obviating the necessity to implement reality conditions.  Hence this eliminated one of the main difficulties associated with the Ashtekar variables, however at the price of a more complicated Hamiltonian constraint than the one based on a complex connection.  Moreover, a further criticism is the observation that by going to a real connection, one loses the interpretation of $A^a_i$ as the pullback to 3-space $\Sigma$ of a spacetime connection, which one had in the self-dual case \cite{SAMUEL}.  Georgio Immirzi observed that the canonical transformation utilized by Barbero could be generalized to include a complex number $\beta$ known as the Immirzi parameter, which generalizes the Ashtekar connection and consequent formalism.  The Immirzi parameter does not affect the classical dynamics of GR, but at the same time the canonical transformation cannot be implemented unitarily in the quantum theory.\footnote{One effect is that the Immirzi parameter rescales the spectra of geometric operators in LQG.}  So one ends up with a theory of connections $A^a_i=\Gamma^a_i+\beta{K}^a_i$, labelled by a free parameter $\beta$ which does not appear to be fixed by any theoretical considerations.\par 
\indent 
Recently, a new development has occurred, in which it was demonstrated that the Wheeler--DeWitt equation with cosmological constant $\Lambda$ could be solved from a group theoretical context in the Ashtekar variables.  In \cite{CIS} it was shown that the quantum Hamiltonian constraint can be written as the Poisson bracket of the imaginary part $Q$ of the Chern--Simons functional of the Ashtekar connection, and the local volume functional $V(x)$, as in
\begin{eqnarray} 
\label{VOLUME}
\{Q,V(x)\}=-\frac{1}{2}G\Lambda{V}(x).
\end{eqnarray}
\noindent 
Quantization of the theory proceeded according the the Algebraic quantization program of Ashtekar, using a set $S=\{I,V(x),Q\}$, consisting of essentially three elements.  Upon the observation that restricted to phase space configurations satisfying the Wheeler--DeWitt equation (\ref{VOLUME}), that Q and $V(x)$ then satisfy the Lie algebra of the affine group of transformations of the straight line $\textbf{aff(R)}$, it became clear that all states of the physical Hilbert space $\textbf{H}_{Phys} $must necessarily come from unitary, irreducible representations of the affine group $Aff(R)$.\par 
\indent
The results of \cite{CIS} were based on the self-dual case $(\beta=i)$.  As it was just the imaginary part $Q$, rather than the full Chern--Simons functional $I_{CS}[A]$ which was needed to obtain hermitian operators, then this addressed the issue of reality conditions.  The value $\beta=i$ was chosen in \cite{CIS} precisely to avoid the complicated form of the Hamiltonian constraint.  In the present paper we will generalize the results of \cite{CIS} in at least two ways: (i) We will generalize the solutions to incorporate arbitrary values of the Immirzi parameter $\beta$.  A motivation for this is that one now has access to a rich calculus on the space of Wilson loops of connections.  This could open up further possibilities for the investigation of consequences of the affine group formalism, as well as make contact with Loop Quantum Gravity (LQG). (ii) The set $S$ consisted of three elements, and therefore does not separate the full classical phase space of GR.  In this paper we will enlarge $S$ to include more elements.  Specifically, we will enlarge $S$ to include the set of gauge invariant, diffeomorphism invariant observables generated by the Chern--Simons and volume functionals, seen as fundamental objects.\par 
\indent 
The organization of this paper is as follows.  In Section 2 we introduce the starting action, namely the Holst action, parametrized by $\beta$, and its 3+1 decomposition.  This corresponds on-shell classically to GR for arbitrary $\beta$.  In Section 3 we collect various Poisson bracket identities, and we re-write the Hamiltonian constraint entirely in terms of Poisson brackets involving the Chern--Simons functional $I_{CS}[A]$ and the volume $V(x)$.  In section 4 we perform a quantization, leaving open the task of constructing elements of the physical Hilbert space $\textbf{H}_{Phys}$ for future research.

\section{Starting action and the constraints}

The fundamental starting point for this paper will be the Holst action \cite{HOLST} based on tetrads $e^I_{\mu}$ and a Lorentz spin connection $A^{IJ}_{\mu}$
\begin{eqnarray} 
\label{HOLST}
S[e,A]=\frac{1}{2}\int_Md^4x~ee^{\mu}_Ie^{\nu}_J\bigl(F_{\mu\nu}^{IJ}-\frac{1}{2\beta}\epsilon^{IJ}_{JK}F^{KL}_{\mu\nu}\bigr),
\end{eqnarray} 
\noindent 
where $e=\sqrt{-g}$ is the determinant of the spacetime metric, $F_{\mu\nu}^{IJ}$ is the curvature of a $SO(3,1)$ connection $A_{\mu}^{IJ}$, and $\beta$ is the Barbero--Immirzi parameter.\footnote{For index conventions, $I,J,\dots$ refer to internal Lorentz indices, while $\mu,\nu,\dots$ are spacetime indices, each taking values $0,1-3$.  We will refer to internal $SU(2)$ indices by $a,b,.\dots$ and spatial indices by $i,j,\dots$, the latter two sets taking values $1-3$.}  Variation of (\ref{HOLST}) with respect to the connection $A^{IJ}_{\mu}$ yields
\begin{eqnarray} 
\label{HOLST1}
D_{\mu}(ee^{\mu}_{[I}e^{\nu}_{J]})=0,
\end{eqnarray} 
\noindent 
where $D_{\mu}$ denotes the covariant derivative acting on both spacetime and Lorentz indices.  This implies 
\begin{eqnarray} 
\label{HOLSTER}
A^{IJ}_{\mu}=e^I_{\nu}\nabla_{\mu}e^{J\nu},
\end{eqnarray} 
\noindent 
namely that on-shell, $A^{IJ}_{\mu}$ is the Levi--Civita spin connection compatible with the tetrads $e^I_{\mu}$.  The term in the action (\ref{HOLST}) proportional to $\frac{1}{\beta}$ then reduces on-shell to 
\begin{eqnarray} 
\label{HOLSTER1}
ee^{\mu}_Ie^{\nu}_J\epsilon^{IJ}_{KL}F^{KL}_{\mu\nu}=\epsilon^{\mu\nu\rho\sigma}R_{\mu\nu\rho\sigma}=0
\end{eqnarray}
\noindent 
since $R_{[\mu\nu\rho]}^{\sigma}=0$ on account of the first Bianchi identity, where $R_{\mu\nu\rho\sigma}$ is the Riemann curvature tensor of the spacetime metric $g_{\mu\nu}=\eta_{IJ}e^I_{\mu}e^J_{\nu}$ constructed from the tetrads.  So one sees that for all values of $\beta$, the action (\ref{HOLST}) is classically the same theory as vacuum Einstein's General Relativity (GR).\par 
\indent 
Given that (\ref{HOLST}) classically is indeed GR for arbitrary $\beta$, one may then proceed with a canonical analysis of the action, yielding (see \cite{HOLST} for the details)
\begin{eqnarray}
\label{HOLST2}
S=\frac{1}{\kappa\beta}\int{dt}\int_{\Sigma}d^3x\bigl(\widetilde{E}^i_a\pounds_tA^a_i-\Lambda^aG_a-N^iH_i-NH\bigr).
\end{eqnarray}
\noindent 
where $\kappa=\frac{8\pi{G}_{Newton}}{c^3}$.  Here, $A^a_i=\Gamma^a_i+\beta{K}^a_i$ is a spatial $SU(2)$ connection, labelled by the Immirzi parameter $\beta$, which forms a canonically conjugate pair with the densitized triad $\widetilde{E}^i_a$.  The quantity $\Gamma^a_i$, the triad compatible connection, is given by \cite{THIEMANN}
\begin{eqnarray} 
\label{COMPATIBLE}
\Gamma^a_i=\frac{1}{2}\epsilon^{abc}E^j_c\bigl(\partial_je^b_i-\partial_ie^b_j+E^k_be^d_i\partial_je^d_j\bigr)
\end{eqnarray} 
\noindent 
where $E^i_ae^a_j=\delta^i_j$, and $E^i_ae^b_i=\delta^b_a$.  The auxiliary fields $\Lambda^a$, $N^i$ and $N$ are Lagrange multiplier fields, which smear their respective constraints.  We have the Gauss' Law, vector and Hamiltonian constraints (we have included a cosmological constant term), given respectively by \cite{HOLST} 
\begin{eqnarray} 
\label{HOLST3}
G_a=D_i\widetilde{E}^i_a=\partial_i\widetilde{E}^i_a+\epsilon_{ab}^cA^b_i\widetilde{E}^i_c;\nonumber\\
H_i=\eps\widetilde{E}^j_a\widetilde{B}^{ka}-\frac{1+\beta^2}{\beta}K^a_iG_a;\nonumber\\
{H}=\frac{\eps\epsilon^{abc}\widetilde{E}^i_a\widetilde{E}^j_b}{2\sqrt{\vert\hbox{det}\widetilde{E}\vert}}\Bigl(\widetilde{B}^k_c+\frac{\Lambda}{3}\widetilde{E}^k_c
-\frac{1}{2}(1+\beta^2)\widetilde{\epsilon}^{kmn}\epsilon_{cde}K^d_mK^e_n\Bigr),
\end{eqnarray} 
\noindent 
where $\widetilde{B}^{ia}=\frac{1}{2}\widetilde{\epsilon}^{ijk}F^a_{jk}$ is the magnetic field of the connection $A^a_i$.  From the canonical structure of (\ref{HOLST2}), one reads off the elementary Poisson brackets
\begin{eqnarray} 
\label{HOLST4}
\{A^a_i(x),\widetilde{E}^j_b(y)\}=\kappa\beta\delta^a_b\delta^j_i\delta^{(3)}(x,y);~~\{K^a_i(x),\widetilde{E}^j_b(y)\}=\kappa\delta^a_b\delta^j_i\delta^{(3)}(x,y).
\end{eqnarray} 
\noindent 
For $\beta=\pm{i}$ we have the self-dual case with complex Ashtekar variables.  This causes the extrinsic curvature squared terms of (\ref{HOLST3}) vanish, considerably simplifying the Hamiltonian constraint.  This was the case considered in \cite{CIS}.  As one can see, the existence of the extrinsic curvature squared terms of the Hamiltonian constraint for $\beta\neq\pm{i}$ seems in general to render the constraint more difficult to implement.  The results of the present paper will show that the Hamiltonian constraint for arbitrary $\beta$ actually is as straightforward as the self-dual $\beta=\pm{i}$ case of \cite{CIS}.\par 
\indent

\section{Poisson bracket manipulations and identities}

Prior to quantization of the theory (\ref{HOLST2}), we will first write each term of the Hamiltonian constraint as a Poisson bracket.  Specifically, we will show that each such term can be reduced entirely and explicitly to Poisson brackets involving just the volume $V(x)$ and the Chern--Simons $I_{CS}[A]$ functionals.  To get the curvature term of the Hamiltonian constraint, also known as the Euclidean Hamiltonian constraint $\mathcal {\cal{H}}_E(x)$, defining $q=\vert\hbox{det}\widetilde{E}(x)\vert$ as the absolute value of the determinant of the densitized triad, one contracts the identity
\begin{eqnarray} 
\label{HAMILTONIAN1}
\{A^a_i(y),\sqrt{{q}(x)}\}=(\kappa\beta)\frac{\eps\epsilon^{abc}\widetilde{E}^i_a\widetilde{E}^j_b}{4\sqrt{{q}}}\delta^{(3)}(x,y)
\end{eqnarray}
\noindent 
with the magnetic field $\widetilde{B}^i_a(y)$ and integrates $\int_{\Sigma}d^3y$, rewriting as a Poisson bracket with the Chern-Simons functional $I_{CS}[A]$ of the connection $A^a_i$
\begin{eqnarray} 
\label{HAMILTONIAN2}
\mathcal {\cal{H}}_E(x)=\frac{\eps\epsilon^{abc}\widetilde{E}^i_a(x)\widetilde{E}^j_b(x)\widetilde{B}^k_c(x)}{2\sqrt{q(x)}}\nonumber\\
=\frac{2}{(\kappa\beta)}\int_{\Sigma}d^3y\widetilde{B}^i_a(y)\{A^a_i(y),\sqrt{q(x)}\}=\frac{2}{(\kappa\beta)}\{I_{CS}[A],\sqrt{q(x)}\}.
\end{eqnarray}
\noindent 
Equation (\ref{HAMILTONIAN2}) is somewhat reminiscent of a trick due to Thiemann \cite{THIEMANN1}, where the Euclidean Hamiltonian constraint can be written as (adapted to the notation of the present paper)
\begin{eqnarray} 
\label{THIEMAN}
{\cal{H}}^E[N]=\frac{2}{\kappa}\int_{\Sigma}d^3xN(x)\widetilde{\epsilon}^{ijk}\hbox{tr}(F_{ij}\{A_k,V\}.
\end{eqnarray} 
\noindent 
The quantization of (\ref{THIEMAN}) then proceeds via LQG techniques, wherein one replaces the connection and curvature with holonomies adapted to triangulations $\Delta$ of a graph corresponding to a spin network state
\begin{eqnarray} 
\label{THIEMAN1}
{\cal{H}}^E_{\Delta}[N]=-\frac{2}{3}N_v\widetilde{\epsilon}^{ijk}\hbox{tr}\bigl(h_{\alpha_{ij}(\Delta)}h_{s_k(\Delta)}\{h^{-1}_{s_k(\Delta)},V\}\bigr).
\end{eqnarray}
\noindent 
But there is one essential difference between (\ref{HAMILTONIAN2}) and (\ref{THIEMAN}).  In (\ref{HAMILTONIAN2}), the curvature $F_{ij}$ has been brought inside the Poisson bracket to yield the Chern--Simons functional, whereas in (\ref{THIEMAN}), it remains outside it.  The Chern--Simons functional is given in the language of differential forms by
\begin{eqnarray} 
\label{CHERNIT}
I_{CS}[A]=\frac{1}{2}\int_{\Sigma}\hbox{tr}\bigl({A}\wedge{dA}+\frac{2}{3}{A}\wedge{A}\wedge{A}\bigr).
\end{eqnarray}
\noindent 
Note, for spatial 3-manifolds $\Sigma$ without boundary or alternatively, for field configurations having sufficiently rapid falloff conditions on the boundary $\partial\Sigma$, that (\ref{CHERNIT}) is a gauge-invariant, spatially diffeomorphism-invariant functional of the connection $A^a_i$.  Equation (\ref{THIEMAN1}) can be seen essentially as an adaptation of the Chern-Simons functional to LQG representations where the volume element and holonomies are fundamental.  In the affine group formalism, we will rather regard the $Chern-Simons$ and $volume$ functionals as fundamental, since we can then exploit their invariances in the construction of physical states.\par 
\indent 
The cosmological constant term of the Hamiltonian constraint is given by
\begin{eqnarray} 
\label{HAMILTONIAN3}
\frac{\Lambda\eps\epsilon^{abc}\widetilde{E}^i_a(x)\widetilde{E}^j_b(x)\widetilde{E}^k_c(x)}{6\sqrt{q(x)}}=\Lambda\sqrt{q(x)},
\end{eqnarray}
\noindent 
which in conjunction with (\ref{HAMILTONIAN2}) forms the basis for the results of \cite{CIS}.\par 
\indent  
The extrinsic curvature squared terms of the Hamiltonian constraint are the terms proportional to $1+\beta^2$, which are no longer zero since $\beta\neq{\pm}i$ in general, and will require some additional care.  To address these, let us consider the following functional of the triadic extrinsic curvature $K^a_i$
\begin{eqnarray} 
\label{HAMILTONIANFOUR}
\int_{\Sigma}d^3y(\hbox{det}K(y))=\frac{1}{3!}\int_{\Sigma}d^3y\widetilde{\epsilon}^{ijk}\epsilon_{abc}K^a_i(y)K^b_j(y)K^c_k(y).
\end{eqnarray}
\noindent 
This satisfies the following Poisson bracket identity
\begin{eqnarray}
\label{HAMILTONIAN4}
\frac{1}{\kappa}\{\int_{\Sigma}d^3y(\hbox{det}K(y)),\widetilde{E}^k_c(x)\}=\frac{1}{2!}\widetilde{\epsilon}^{kmn}\epsilon_{cde}K^d_m(x)K^e_n(x).
\end{eqnarray}
\noindent 
To obtain the extrinsic curvature squared contribution to the Hamiltonian constraint, the term proportional to $1+\beta^2$, let us contract (\ref{HAMILTONIAN4}) with two factors of $\widetilde{E}^i_a$ and perform the following manipulations
\begin{eqnarray} 
\label{HAMILTONIAN5}
\frac{\eps\epsilon^{abc}\widetilde{E}^i_a(x)\widetilde{E}^j_b(x)}{2\kappa\sqrt{q(x)}}\{\int_{\Sigma}d^3y(\hbox{det}K(y)),\widetilde{E}^k_c(x)\}\nonumber\\
=\frac{\eps\epsilon^{abc}\widetilde{E}^i_a(x)\widetilde{E}^j_b(x)}{4\sqrt{q(x)}}\widetilde{\epsilon}^{kmn}\epsilon_{cde}K^d_m(x)K^e_n(x)\nonumber\\
=\frac{2}{\kappa}\{\int_{\Sigma}d^3y(\hbox{det}K(y)),\sqrt{q(x)}\}.
\end{eqnarray} 
\noindent
In analogy to the curvature term, in (\ref{HAMILTONIAN5}) we have brought all quantities inside the Poisson bracket.  Defining the local volume function $V(x)$, given by
\begin{eqnarray}
\label{LOCALVOLUME} 
V(x)=\sqrt{q(x)}=\sqrt{\Bigl\vert\frac{1}{6}\eps\epsilon^{abc}\widetilde{E}^i_a(x)\widetilde{E}^j_b(x)\widetilde{E}^k_c(x)\Bigr\vert},
\end{eqnarray}
\noindent 
then one sees that the Hamiltonian constraint, the third equation of (\ref{HOLST3}), can be written as
\begin{eqnarray} 
\label{HAMILTONIAN6}
\frac{2}{\beta\kappa}\{I_{CS}[A],V(x)\}+\frac{2(1+\beta^2)}{\kappa}\{\int_{\Sigma}d^3y(\hbox{det}K(y)),V(x)\}+\Lambda{V}(x)=0.
\end{eqnarray}
\noindent 
Having written the Hamiltonian constraint as a Poisson bracket (\ref{HAMILTONIAN6}), one sees the ingredients in place for an affine algebraic structure of the sum of the Chern--Simons functional and the integrated determinant of the extrinsic curvature with the local volume operator functional $V(x)$.  One could attempt to perform an affine group quantization in accordance with \cite{CIS}.  However, the extrinsic curvature $(\hbox{det}K)$ term appears ostensibly to be new.  Indeed, the treatment of this so-called kinetic term due to Thiemann is given by \cite{THIEMANN1}
\begin{eqnarray} 
\label{NOWTHIEMANN}
T[N]=8\int_{\Sigma}d^3x\frac{N}{\kappa^3}\widetilde{\epsilon}^{ijk}\hbox{tr}\bigl(\{A_i,{\cal{K}}\}\{A_j,{\cal{K}}\}\{A_k,V\}\bigr),
\end{eqnarray} 
\noindent
with ${\cal{K}}$ given by (\ref{EXTRINSIC}), wherein one eliminates all occurences of the connection in favor of holonomies and reduces ${\cal{K}}$ accordingly prior to quantization.  We will adopt an alternate approach adapted to the affine group representation formalism, where it is the Chern--Simons functional $I_{CS}[A]$ and the volume $V$ which are the central geometric entities.  Therefore we will reduce the kinetic $(\hbox{det}K)$ term accordingly to Poisson brackets involving just these quantities prior to quantization of the theory.

\subsection{The kinetic term of the Hamiltonian constraint}

We will now carry out the decomposition of the $(\hbox{det}K)$ term in the Hamiltonian constraint as previously noted, writing it completely in terms of Poisson brackets involving $I_{CS}[A]$ and $V(x)$.  To do so let us define the following functional $\mathcal {\cal{K}}$ for the densitized trace of the extrinsic curvature of $\Sigma$, and the (global) volume $V$ by
\begin{eqnarray} 
\label{EXTRINSIC}
\mathcal {\cal{K}}=\int_{\Sigma}d^3xK^a_i\widetilde{E}^i_a;~~V=\int_{\Sigma}d^3xV(x).
\end{eqnarray} 
\noindent 
Note that $\mathcal {\cal{K}}$ in (\ref{EXTRINSIC}) can be written completely in terms of Poisson brackets involving $I_{CS}[A]$ and $V$ via
\begin{eqnarray} 
\label{EXTRAINSIC}
\mathcal {\cal{K}}=\frac{1}{(\beta^2\kappa)}\{\mathcal {\cal{H}}_{E},V\}=\frac{2}{(\beta^3\kappa^2)}\big\{\{I_{CS}[A],V\},V\big\},
\end{eqnarray} 
\noindent 
where $\mathcal {\cal{H}}_E=\int_{\Sigma}d^3x\mathcal {\cal{H}}_E(x)$ is the (unsmeared) integral of the Euclidean Hamiltonian constraint.\par 
\indent  
Define the one form $k^a=K^a_idx^i$, and consider the Taylor expansion of the Chern--Simons functional
\begin{eqnarray} 
\label{EXTRINSIC1}
I_{CS}[A]=I_{CS}[\Gamma+\beta{k}]
=I_{CS}[\Gamma]+\beta\int_{\Sigma}{k^a}\wedge{R}^a[\Gamma]\nonumber\\
+\frac{\beta^2}{2!}\int_{\Sigma}{k^a}\wedge{D^{\Gamma}k^a}+\beta^3\int_{\Sigma}d^3x(\hbox{det}K),
\end{eqnarray} 
\noindent 
where $\widetilde{R}^{ia}[\Gamma]=\widetilde{\epsilon}^{ijk}\partial_j\Gamma^a_k+\frac{1}{2}\widetilde{\epsilon}^{ijk}\epsilon^a_{bc}\Gamma^b_j\Gamma^c_k$ is the magnetic field corresponding to the connection $\Gamma^a_i$, which depends on the triad $e^a_i$.  The Poisson bracket of (\ref{EXTRINSIC1}) with $V(x)$ is given by
\begin{eqnarray} 
\label{EXTRINSIC22}
\{I_{CS}[A],V(x)\}=\beta\{\int_{\Sigma}{k^a}\wedge{R}^a[\Gamma],V(x)\}+\beta^3\{\int_{\Sigma}d^3y(\hbox{det}K),V(x)\}.
\end{eqnarray}
\noindent 
To get (\ref{EXTRINSIC22}) we have used $\{I_{CS}[\Gamma],V(x)\}=0$ since $\Gamma^a_i=\Gamma^a_i[e]$ depends just on the triad (\ref{COMPATIBLE}) and therefore Poisson commutes with $V(x)$.  Additionally, the term of order $\beta^2$ in (\ref{EXTRINSIC22}) has a vanishing Poisson bracket  
\begin{eqnarray} 
\label{EXTRINSIC33}
\frac{1}{2!}\{\int_{\Sigma}{k^a}\wedge{D^{\Gamma}k^a},V(x)\}=\frac{1}{2!}\{\int_{\Sigma}d^3x\widetilde{\epsilon}^{ijk}K^a_iD^{\Gamma}_jK^a_k,V(x)\}\nonumber\\
=(\beta\kappa)\widetilde{\epsilon}^{ijk}e^a_iD^{\Gamma}_jK^a_k=D^{\Gamma}_j(\widetilde{\epsilon}^{ijk}e^a_iK^a_k)=0
\end{eqnarray} 
\noindent 
due to the Gauss' law constraint of ADM triad variables, combined with the fact that $\Gamma^a_i$ is the torsion-free connection compatible with the (undensitized) triad $e^a_i$.  We see in (\ref{EXTRINSIC22}) that the extrinsic curvature squared term of the Hamiltonian constraint, if not for the term linear in $\beta$, can be written almost entirely as a Poisson bracket involving just $I_{CS}[A]$ and $V(x)$.  We will need a second identity involving this term linear in $\beta$, so that it can be eliminated in favor of these fundamental objects.  This brings into play the following identity involving the Ashtekar magnetic field
\begin{eqnarray} 
\label{EXTRINSIC4}
\widetilde{B}^{ia}[A]=\widetilde{B}[\Gamma+\beta{k}]
=\widetilde{R}^{ia}[\Gamma]+\beta\widetilde{\epsilon}^{ijk}D^{\Gamma}_jK^a_k+\frac{\beta^2}{2}\widetilde{\epsilon}^{ijk}\epsilon^{abc}K^b_jK^c_k.
\end{eqnarray}
\noindent 
To put (\ref{EXTRINSIC4}) on an equal footing with (\ref{EXTRINSIC1}) let us contract (\ref{EXTRINSIC4}) with $K^a_i$ and integrate over $\Sigma$.  This yields
\begin{eqnarray} 
\label{EXTRINSIC5}
\beta\int_{\Sigma}d^3x\widetilde{B}^i_aK^a_i=\beta\int_{\Sigma}{k^a}\wedge{R^a[\Gamma]}+\beta^2\int_{\Sigma}{k^a}\wedge{D^{\Gamma}k^a}+
3\beta^3\int_{\Sigma}d^3x(\hbox{det}K).
\end{eqnarray}
\noindent 
We will also need to write the left hand side of (\ref{EXTRINSIC5}) as a Poisson bracket completely in terms of $I_{CS}[A]$ and $V$ but first, let us take the Poisson bracket of (\ref{EXTRINSIC5}) with $V(x)$ in preparation of the ingredients for the Hamiltonian constraint.  The second term, of order $\beta^2$, has a vanishing Poisson bracket as shown in (\ref{EXTRINSIC33}).  So we are left with
\begin{eqnarray} 
\label{EXTRINSIC6}
\{\beta\int_{\Sigma}d^3y\widetilde{B}^i_aK^a_i,V(x)\}=\beta\{\int_{\Sigma}{k^a}\wedge{R^a[\Gamma]},V(x)\}+3\beta^3\{\int_{\Sigma}d^3x(\hbox{det}K),V(x)\}.
\end{eqnarray}
\noindent 
Since $\mathcal {\cal{K}}$ is the generator of constant scale transformations of the basic variables, for instance one has
\begin{eqnarray} 
\label{KAYY}
\{\mathcal {\cal{K}},K^a_i(x)\}=-{\kappa}K^a_i(x);~~\{\mathcal {\cal{K}},V(x)\}=\frac{3\kappa}{2}V(x),
\end{eqnarray} 
\noindent
then the following identity holds
\begin{eqnarray} 
\label{POISSON}
\frac{1}{(\beta\kappa)}\{I_{CS}[A],\mathcal {\cal{K}}\}
=-\frac{1}{(\beta\kappa)}\int_{\Sigma}d^3x\widetilde{B}^i_a\{\mathcal {\cal{K}},A^a_i\}=\int_{\Sigma}d^3x\widetilde{B}^i_aK^a_i.
\end{eqnarray}
\noindent 
That (\ref{POISSON}) holds follows from the observation that $A^a_i=\Gamma^a_i+\beta{K}^a_i$ and $\Gamma^a_i$, being a homogeneous rational function of order zero in the triads and their derivatives (\ref{COMPATIBLE}), is invariant under constant scale transformations \cite{THIEMANN}.  Then using (\ref{EXTRAINSIC}), one sees that the right hand side of (\ref{POISSON}) can be written in terms of Poisson brackets involving just $I_{CS}[A]$ and $V$ as
\begin{eqnarray} 
\label{POISSON1}
\beta\int_{\Sigma}d^3x\widetilde{B}^i_aK^a_i
=\frac{2}{(\beta\kappa)^3}\Big\{I_{CS}[A],\big\{\{I_{CS}[A],V\},V\big\}\Big\}.
\end{eqnarray} 
\noindent 
Taking (\ref{EXTRINSIC22}), and the substitution of (\ref{POISSON1}) into (\ref{EXTRINSIC6}) we have the following system of simultaneous equations (i) and (ii)
\begin{eqnarray}
\label{EXTRINSIC2}
(i)~~\{I_{CS}[A],V(x)\}=\beta\{\int_{\Sigma}{k^a}\wedge{R}^a[\Gamma],V(x)\}+\beta^3\{\int_{\Sigma}d^3y(\hbox{det}K),V(x)\};\nonumber\\
(ii)~~\frac{2}{(\beta\kappa)^3}\Big\{\big\{I_{CS}[A],\{\{I_{CS}[A],V\},V\}\big\},V(x)\Big\}\nonumber\\
=\beta\{\int_{\Sigma}{k^a}\wedge{R^a[\Gamma]},V(x)\}+3\beta^3\{\int_{\Sigma}d^3x(\hbox{det}K),V(x)\}.
\end{eqnarray}
\noindent 
Seen as a simultaneous linear system of two equations in two unknowns, the unknowns being the coefficients of $\beta$ and $\beta^3$, the left hand sides have been written explicitly and entirely in terms of Poisson brackets involving the Chern--Simons functional $I_{CS}[A]$ and the volume operator $V$.  The solution to this system is given by
\begin{eqnarray} 
\label{EXTRINSIC3}
\{\int_{\Sigma}{k^a}\wedge{R^a[\Gamma]},V(x)\}=\frac{1}{2\beta}\biggl[3\{I_{CS}[A],V(x)\}\nonumber\\
-\frac{2}{(\beta\kappa)^3}\Big\{\big\{I_{CS}[A],\{\{I_{CS}[A],V\},V\}\big\},V(x)\Big\}\biggr];\nonumber\\
\{\int_{\Sigma}d^3y(\hbox{det}K(y)),V(x)\}=\frac{1}{2\beta^3}\biggl[-\{I_{CS}[A],V(x)\}\nonumber\\
+\frac{2}{(\beta\kappa)^3}\Big\{\big\{I_{CS}[A],\{\{I_{CS}[A],V\},V\}\big\},V(x)\Big\}\biggr].
\end{eqnarray}
\noindent 
The second equation of (\ref{EXTRINSIC3}) is the sought-after relation which recasts the extrinsic curvature squared term of the Hamiltonian constraint, through its Poisson bracket in terms of $(\hbox{det}K)$, as a Poisson bracket involving just the fundamental entities $I_{CS}[A]$ and $V(x)$.  The first equation involving ${k^a}\wedge{R^a[\Gamma]}$ is so analogously expressed, and while having some significance, is not needed for the purposes of re-writing the Hamiltonian constraint.

\section{The Hamiltonian constraint: revisited}

Having obtained the desired relation in the second equation of (\ref{EXTRINSIC3}), we will now substitute this back into the Hamiltonian constraint (\ref{HAMILTONIAN6}).  So multiplying (\ref{HAMILTONIAN6}) by $(\beta\kappa)/2$, we obtain the Hamiltonian constraint for arbitrary value of the Immirzi parameter $\beta$ as
\begin{eqnarray} 
\label{REALHAMILTONIAN}
\Bigl(1-\frac{1+\beta^2}{2\beta^2}\Bigr)\{I_{CS}[A],V(x)\}\nonumber\\
+\Bigl(\frac{1+\beta^2}{\beta^5\kappa^3}\Bigr)\Big\{\big\{I_{CS}[A],\big\{I_{CS}[A],V\},V\}\big\},V(x)\Big\}+\frac{(\beta\kappa\Lambda)}{2}V(x)=0.
\end{eqnarray}
\noindent 
As claimed, this has been completely reduced to Poisson brackets involving $V(x)$ and $I_{CS}[A]$, seen as fundamental geometric objects for gravity.  We are now ready to proceed with quantization.  But first, let us comment on an issue stemming from \cite{CIS}, related to algebraic quantization.\par 
\indent  
The quantization in \cite{CIS} originated in the existence of a certain set $S=(I,V(x),Q=Im[I_{CS}[A]])$, which was closed under Poisson brackets and under complex conjugation, which brings up three issues: (i) While the local Hamiltonian constraint could be written explicitly and entirely in terms of the objects of $S$ at a classical level, the requirement of closure under Poisson brackets was tantamount to implementation of the Hamiltonian constraint at the classical level prior to quantization. (ii) The set S, consisting essentially of just three elements, does not separate the full classical phase space of General relativity.  So the quantization of \cite{CIS} pertained to just the Wheeler--DeWitt equation, which had already been solved at the classical level, and not to the full classical phase space of GR. (iii) While quantization of the full classical phase space of GR was not necessary in order to construct certain elements of the physical Hilbert space $\textbf{H}_{Phys}$, it is still worthwhile to ask to what extent the set $S$ may be enlarged prior to quantization.\par 
\indent 
In this paper we will implement the Hamiltonian constraint subsequent (and not prior to) to quantization, by promoting Poisson brackets to $\frac{1}{i\hbar}$ time commutators subsequent to defining the set $S$.  Consider the following gauge-invariant, spatially diffeomorphism invariant ($SU(2)*diff(\Sigma)$) objects $I$, $V$ and $I_{CS}[A]$ and define $S_0$ as the set consisting of these objects along with all possible Poisson brackets generated from them.  So $\forall{A},B\in{S}_0$, we have that $\{A,B\}\in{S}_0$.  This follows from the Jacobi identity for Poisson brackets.  Let $T[u]$ be the generator of the symmetry, where $\{A,T[u]\}=\{B,T[u]\}=0$ for invariant functions $A$ and $B$.  Then from the Jacobi identity we have
\begin{eqnarray} 
\label{JACOBI}
\big\{\{A,T[u]\},B\big\}+\big\{\{T[u],B\},A\big\}+\big\{\{A,B\},T[u]\big\}=0,
\end{eqnarray} 
\noindent 
which implies that $\{A,B\}$ is also $SU(2)*diff(\Sigma)$ invariant.  So the algebra of observables constructed from $I_{CS}[A]$ and $V$ is closed in the sense that it is $SU(2)*diff(\Sigma)$ invariant.  Note that $S_0$ will include elements of the form
\begin{eqnarray} 
\label{OFFORM}
\{I_{CS}[A],V\},~~\big\{\{I_{CS}[A],V\},V\big\},~~\Big\{I_{CS}[A],\big\{\{I_{CS}[A],V\},V\big\}\Big\},
\end{eqnarray}
\noindent 
which are all ingredients necessary to form the Hamiltonian constraint (\ref{REALHAMILTONIAN}).\par 
\indent
The set $S_0$ consists completely of $SU(2)*diff(\Sigma)$ invariant objects, which are globally defined, and hence is not large enough to produce the local Hamiltonian constraint $H(x)$.  So let us define the set $S=S_0\cup{V}(x)$, by appending the local volume functional, and all Poisson brackets thereof such that $\forall{A},B\in{S}$, we have $\{A,B\}\in{S}$.  So $S$ is closed under Poisson brackets and under complex conjugation (since all elements are real for real $\beta$), and contains the ingredients of the local Hamiltonian constraint $H(x)=0$.  Additionally, $S$ has an infinite number of elements, and contains a large class of gauge invariant, diffeomorphism invariant observables as a subset.\footnote{We have enlarged the set $S$ in relation to \cite{CIS}, but we have not shown whether or not $S$ separates the points of the full classical phase space of GR.}  For the purposes of the present paper, we will use $S$ as the starting point for quantization.\par 
\indent 

\subsection{Quantization}

We will quantize the Hamiltonian constraint, noting from group theoretical considerations that its kernel is nonempty.  We can now proceed with quantization by promotion of all Poisson brackets in (\ref{REALHAMILTONIAN}) to $\frac{1}{(i\hbar)}$ times quantum commutators.  The condition that a state $\vert\psi\rangle\in{Ker}H(x)$ be annihilated by the local quantum Hamiltonian constraint is given by
\begin{eqnarray} 
\label{REALHAMILTONIAN1}
\frac{1}{2}\Bigl(1-\frac{1}{\beta^2}\Bigr)\frac{1}{(i\hbar)}\bigl[\widehat{I}_{CS}[A],\widehat{V}(x)\bigr]\vert\psi\rangle\nonumber\\
+\Bigl(\frac{1+\beta^2}{\beta^5\kappa^3}\Bigr)\Bigl(\frac{1}{i\hbar}\Bigr)^4\bigl[\bigl[\widehat{I}_{CS}[A],\bigl[\bigl[\widehat{I}_{CS}[A],\widehat{V}\bigr],\widehat{V}\bigr]\bigr],\widehat{V}(x)\bigr]\vert\psi\rangle\nonumber\\
=-\frac{(\beta\kappa\Lambda)}{2}\widehat{V}(x)\vert\psi\rangle.
\end{eqnarray}
\noindent 
Of note is the observation that since $I_{CS}[A]$ and $V(x)$ are composed entirely in terms of connection $A^a_i$ and densitized triad $\widetilde{E}^i_a$ variables respectively, then they are free of ordering ambiguities.  The quantum Hamiltonian constraint (\ref{REALHAMILTONIAN1}) is constructed completely from Poisson brackets involving these quantities.\par 
\indent
It is the main proposition of this paper that (\ref{REALHAMILTONIAN1}) provides an alternative to the LQG quantization of (\ref{THIEMAN}) combined with (\ref{NOWTHIEMANN}).  Let us define the operator $\widehat{O}$ by
\begin{eqnarray} 
\label{REALHAMILTONIAN2}
\widehat{O}=\bigl[\widehat{I}_{CS}[A],\bigl[\bigl[\widehat{I}_{CS}[A],\widehat{V}\bigr],\widehat{V}\bigr]\bigr]
\end{eqnarray}
\noindent 
and $\beta$-dependent constants $p$ and $q$ by
\begin{eqnarray} 
\label{CONSTANTS}
p(\beta)=\frac{1}{2}\Bigl(1-\frac{1}{\beta^2}\Bigr)\frac{1}{(i\hbar)};~~
q(\beta)=\Bigl(\frac{1+\beta^2}{\beta^5\kappa^3}\Bigr)\Bigl(\frac{1}{i\hbar}\Bigr)^4.
\end{eqnarray}
\noindent
Then the quantum Hamiltonian constraint (\ref{REALHAMILTONIAN1}) is given by
\begin{eqnarray} 
\label{REALHAMILTONIAN3}
\bigl[p\widehat{I}_{CS}[A]+q\widehat{O},\widehat{V}(x)\bigr]\vert\psi_{(\beta)}\rangle=-\frac{1}{2}(\beta\kappa\Lambda)\widehat{V}(x)\vert\psi_{(\beta)}\rangle.
\end{eqnarray}
\noindent 
For each value of the Immirzi parameter $\beta$ the states satisfy the affine Lie algebra formed by the generators $Q(\beta)=p\widehat{I}_{CS}[A]+q\widehat{O}$ and the volume operator $\widehat{V}(x)$.  The operator $\widehat{Q}(\beta)$ acts as the dilation operator and the operator $\widehat{V}(x)$ is the dilated operator.  Thus the positivity of the spectrum of $V(x)$ becomes implemented at a quantum level in accordance with the representation $\pi^{+}$.  So we have 
\begin{eqnarray} 
\label{HAMILTONIAN8}
[\widehat{Q}(\beta),\widehat{V}(x)]\vert\psi_{(\beta)}\rangle=-\frac{i(\beta\hbar{G}\Lambda)}{2}\widehat{V}(x)\vert\psi_{(\beta)}\rangle.
\end{eqnarray}
\noindent 
The remaining steps of quantization from (\ref{HAMILTONIAN8}), which involve the actual construction of elements of the physical Hilbert space for arbitrary $\beta$ satisfying the constraints, we will relegate as an open question for future research.

\section{Conclusion}

The main results of this paper have been to demonstrate the relevance of the affine group representation formalism for four dimensional General Relativity.  We have shown that the Hamiltonian constraint with $\Lambda$ term for arbitrary $\beta$ can be reduced completely in terms of Poisson brackets involving the Chern--Simons and volume functional $I_{CS}[A]$ and $V(x)$ and have performed a quantization.  A remaining step is the construction of elements of the physical Hilbert space for arbitrary $\beta$, a task relegated for future research.  Additionally, we have enlarged the set $S$ in relation to \cite{CIS} to include the set of $SU(2)*diff(\Sigma)$ observables generated from $I_{CS}$ and $V$. 

\section{Acknowledgement}

The author is grateful to Chou Ching-Yi for a careful read of the manuscript and for the rectification of various equations.  A full appreciation of the utility of extrinsic curvature-triad variables has been gained.


\begin{thebibliography}{99}

\bibitem{DEWITT} {Bryce S. DeWitt. Phys. Rev. 160, 1113-1148 (1967)}

\bibitem{ASH} {Abhay Ashtekar. Phys. Rev. D36, 1587 (1987)}

\bibitem{ASH1} {Abhay Ashtekar. Phys. Rev. Lett. 57, 2244 (1986)}

\bibitem{BARBERO} {Fernando Barbero. Phys. Rev. D51, 5498 (1995)}

\bibitem{BARBERO1} Barbero F 1995 {\it Phys.\ Rev.} D {\bf 51} 5507-10

\bibitem{SAMUEL} {Joseph Samuel. Class. Quantum Grav. 17: L141-L148, 2000}

\bibitem{IMMIRZI} {Georgio Immirzi. Class. Quantum Grav. 14, 177 (1997)}

\bibitem{CIS}{Chou Ching-Yi, Eyo Ita and Chopin Soo. {\bf 30} (2013): 065013}

\bibitem{HOLST} Holst S 1996 {\it Phys.\ Rev.} D {\bf 53} 5966-69

\bibitem{HOLST3} Corichi A and Reyes J D 2012 {\it J.\ Phys.\ : Conf.\ Ser.} {\bf 360} 012021

\bibitem{THIEMANN} {Thomas Thiemann `Modern Canonical Quantum General Relativity' Cambridge Monographs on Mathematical Physics, 2007}

\bibitem{THIEMANN1} {Thomas Thiemann. Class. Quantum Grav. 15, 839-873 (1998)}












\end{thebibliography}
\end{document}